\documentclass[aps, superscriptaddress ,twocolumn,prl,footinbib]{revtex4-1}
\usepackage{amsmath}
\usepackage{amssymb}
\usepackage{epsfig}
\usepackage{graphics}
\usepackage{bm}
\usepackage{xcolor}

\newcommand{\be}{\begin{equation}}
\newcommand{\ee}{\end{equation}}
\newcommand{\bea}{\begin{eqnarray}}
\newcommand{\eea}{\end{eqnarray}}

\begin{document}
\title{Few-body precursor of the Higgs mode in a  Fermi gas}
\author{J.\ Bjerlin}

\affiliation{Mathematical Physics, LTH, Lund University, SE-22100 Lund, Sweden}

\author{S.\ M.\ Reimann}

\affiliation{Mathematical Physics, LTH, Lund University, SE-22100 Lund, Sweden}

\author{G.\ M.\ Bruun}
\affiliation{Department of Physics and Astronomy, University of Aarhus, Ny Munkegade, DK-8000 Aarhus C, Denmark}

\date{\today}

\begin{abstract}
We demonstrate that an undamped few-body precursor of the Higgs mode can be investigated in a  
harmonically trapped Fermi gas. Using exact diagonalisation, the lowest monopole 
mode frequency is shown to depend non-monotonically on the interaction strength, having 
  a minimum in a crossover region. The minimum deepens with increasing particle number, reflecting that the mode 
  is the few-body analogue of a many-body Higgs mode in the superfluid phase, which has a vanishing frequency at the quantum phase transition
   point to the normal phase.
   We show that this mode   mainly consists of coherent excitations of 
  time-reversed pairs, and that it can be selectively excited by modulating the interaction strength, using for instance  a
  Feshbach resonance in cold atomic gases.  
\end{abstract}

\maketitle
The transition from few-body quantum physics to the thermodynamic limit with increasing particle number is a fundamental problem in science.
A systematic investigation of this question is complicated by the fact that the few-body systems existing in nature, such as atoms and nuclei, 
have limited tunability. Artificially created clusters~\cite{deHeer1993,brack1993} or semiconductor  quantum dots~\cite{reimann2002} offer more flexibility, but they are often strongly coupled to their surroundings making a  study of pure quantum states difficult. 
The creation of highly controllable few-fermion systems using cold atoms in microtraps~\cite{Serwane2011,Zurn2013}, however, 
has opened new perspectives. Tunneling experiments in the few-body limit demonstrated single-atom control~\cite{zurn2012,Rontani2012}.
One has already observed the formation of a Fermi sea~\cite{Wenz2013}, 
 as well as pair correlations in one-dimensional (1D)   few-body atomic gases~\cite{Zurn2013} that have 
 also been studied extensively theoretically~\cite{Sowinski2015a, Sowinski2015b, Grining2015a, Grining2015b,DAmico2015}.
 The few- to many-body transition is arguably  even more interesting in higher dimensions, where quantum phase transitions
  with varying degrees of broken symmetry are ubiquitous~\cite{Sachdevbook}.  A key question 
 concerns the few-body fate of the order parameter, which describes a broken symmetry phase in the thermodynamic limit. 
 
 Another fundamental problem concerns the properties of the Higgs mode, which corresponds to oscillations in the size of the 
 order parameter for a given broken symmetry phase~\cite{Goldtone1961,Higgs1964}. Elementary  particles acquire their mass from the 
 presence of a Higgs mode~\cite{Ryder1996Quantum}, which was famously observed  
 at CERN~\cite{CMS2012,ATLAS2012}. The Higgs mode also leads to collective modes in condensed matter and nuclear 
 systems~\cite{Sachdevbook,BohrMottelson}. Despite its fundamental importance, the list of table top systems 
 where  it has been  observed  is short, mainly because it is typically strongly damped, 
 and because it couples only weakly to experimental probes~\cite{Pekker2015,Cea2014,Cea2015}. Experimental 
 evidence for the existence of a Higgs mode has been reported in disordered and  niobium selenide superconductors~\cite{Sherman2015,Sooryakumar1980,Measson2014,Littlewood1981}.
   Also,      neutron scattering experiments for a quantum anti-ferromagnet~\cite{Ruegg2008} are consistent with the presence of a broad Higgs 
  mode, and  lattice  experiments combined with theoretical models for bosonic atoms in an optical lattice, indicate that a threshold 
  feature  can be interpreted in terms of a strongly damped   Higgs mode~\cite{Endres2012,Liu2015}.

Here, we show how one can explore both these fundamental questions, 
the few- to many-body transition and the nature of the Higgs mode, using an atomic Fermi gas in a new generation of 
 microtraps. We calculate the few-body spectrum 
using exact diagonalisation and show that for closed-shell configurations, the lowest monopole excitation energy depends non-monotonically 
on the interaction strength, having a minimum in a cross-over region, which deepens with 
increasing particle number. By comparing with a many-body theory, we demonstrate that the
mode is the few-body precursor of the Higgs mode in the superfluid phase, which exhibits a vanishing frequency at a quantum phase transition  to 
a normal phase. The  mode mainly consists of time-reversed pair excitations, and it
 can  be selectively excited  by modulating the interaction strength.  

We consider $N/2$  fermions of mass $m$ in each of two 
hyperfine (spin) states $\sigma=\uparrow,\downarrow$ in a 2D harmonic trap $m\omega^2r^2/2$. Particles with opposite spin 
interact via  an attractive delta function interaction (suitably regularised, see below) $g\delta({\mathbf r}-{\mathbf r}')$ with $g<0$, whereas 
particles of the same spin  do not interact.  The Hamiltonian is
\begin{equation}
\hat H=\sum_{i=1}^{N}\left(-\frac{\hbar^2\nabla_i^2}{2m}+\frac12m\omega^2{\mathbf r}_i^2\right)+g\sum_{k,l}\delta({\mathbf r}_k-{\mathbf r}_l)
\label{Hamiltonian}
\end{equation}
where ${\mathbf r}_i=(x_i,y_i)$ is the spatial coordinate of particle $i$, $\nabla_i^2=\partial_{x_i}^2+\partial_{y_i}^2$, and $k$ and $l$ 
in the second sum denote particles with spin $\uparrow $ and spin $\downarrow $, respectively. 

In order to make rigorous predictions unbiased by any assumptions,  
we calculate  the  eigenstates of (\ref{Hamiltonian})  by exact diagonalisation 
using a basis of harmonic oscillator states 
with energy $(2n+ \left| m \right| + 1)\hbar\omega$, where $n=0, 1, 2, 3, \dots $, and $m=0,\pm1,\pm2\ldots$ is the angular momentum.
This method has  been extensively applied to attractive fermion systems, 
mainly in 1D ~\cite{Sowinski2015a, Sowinski2015b, Grining2015a, Grining2015b,DAmico2015} 
but also in 2D~\cite{Rontani2008,Rontani2009}.
As explained in the Supplementary Material~\footnote{See Supplemental Material [url], which includes Refs.~\cite{Bruun2002b,Heiselberg2002}
 for details the theory and for more results. }, we employ a two-parameter cut-off scheme for the basis states  in order to reach maximum convergence.  
Using a sparse representation of the resulting matrix, we find the eigenvectors using the implicitly restarted Arnoldi iteration method~\cite{arpack}.
This generally allows for a significantly larger number of basis states, $\sim10^7$, as compared to other available diagonalisation methods, which 
 is crucial, since we need a very large basis set for an accurate calculation  of 
 the  low-lying collective modes. 

As it stands, the spectrum of $\hat H$ depends logarithmically on the energy cut-off $E_{\text{cut}}$. 
To cure this UV divergence,  we eliminate the coupling constant $g$ and cut-off $E_{\text{cut}}$ in favour of the 
two-body binding energy $\epsilon_b$ per particle. This is defined as $E_2=2\hbar\omega-2\epsilon_b$, where $E_2$ is the ground 
state energy of one $\uparrow$- and 
one $\downarrow$-particle in the trap. In practice, we calculate  $\epsilon_b$ and the many-body spectrum as a function of $g$ 
for the same $E_{\text{cut}}$, and then we plot the spectrum as a function of $\epsilon_b$. Since the two-body problem contains the same 
logarithmic divergence as the many-body problem,  this procedure yields a well-defined theory  for 
$E_{\text{cut}}\rightarrow\infty$~\cite{Randeria1990,Zollner2011,Rontani2008}. A similar UV divergence appears for the system in 3D, where it has been 
regularised using a variety of methods~\cite{Galitskii1958,Gorkov1961,Leggett1980,Bruun1999,Stetcu2007,Alhassid2008,Zinner2009}.

\begin{figure}
\includegraphics[width=0.7\columnwidth]{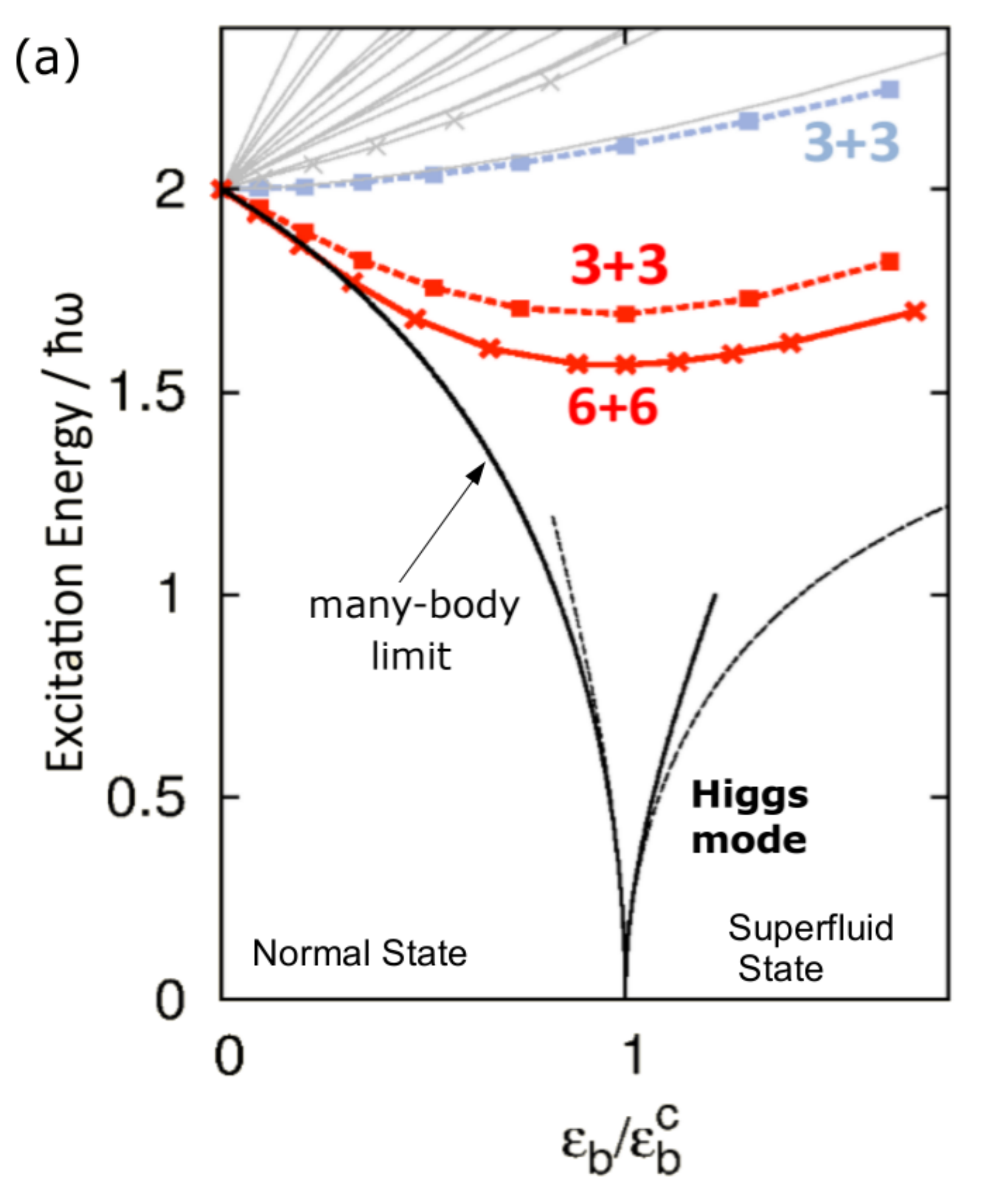}
\caption{\label{SpectrumMagicFig} The lowest monopole excitation
for $3+3$ fermions  (dashed red line) and for $6+6$ fermions (red solid line) obtained by numerical diagonalisation of Eq.\ (\ref{Hamiltonian}). 
The blue dashed line is the second excited state, and the gray solid lines are higher excitations for the $3+3$  system.
The black solid (dotted) lines show the numerical (analytic) many-body 
Higgs-mode energy~\cite{BruunHiggs2014}  (see Supplementary Material~\cite{Note1}). 
}
\end{figure}

\begin{figure}
\includegraphics[width=0.9\columnwidth]{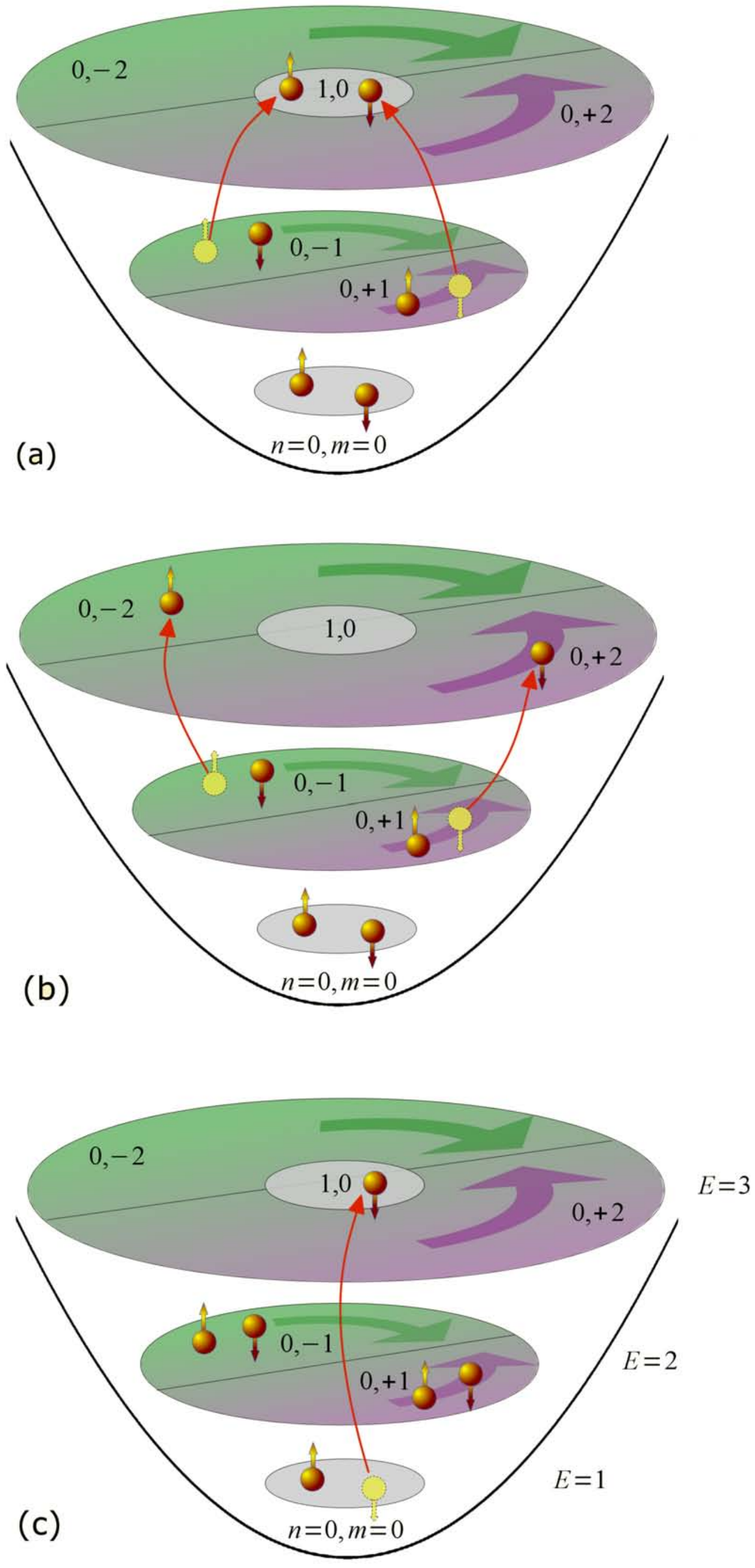}
\caption{\label{TimereverseFig} Panels (a) and (b) show a schematic sketch of an excitation 
of a time-reversed pair $(n,m,\uparrow)$ and $(n,-m,\downarrow)$ one shell up. The energy of such excitations decreases with increasing attraction. 
Panel (c) shows an example of a single-particle monopole excitation two shells up. The energy 
of such excitations grows with increasing attraction.
} 
\end{figure}

Figure \ref{SpectrumMagicFig} shows the lowest monopole (zero  angular momentum) 
excitation spectrum  as a function of the two-body binding energy 
 $\epsilon_b$ for a 3+3 system, consisting of three $\uparrow$-particles and three  $\downarrow$-particles. The non-interacting ground state is 
  a closed-shell configuration with the two lowest harmonic oscillator shells  filled. 
 For no interaction, the  excitations all have the energy $2\hbar\omega$, and they are formed either by 
pair excitations taking 
 two  particles with opposite angular momenta one shell up, see Fig.\ \ref{TimereverseFig}(a)-(b), or by 
 single particle excitations taking  one particle two shells up, see Fig.\ \ref{TimereverseFig}(c). 
 We see that all  excitation energies, except the lowest, increase with increasing attraction since the attractive 
  mean-field interaction potential increases the effective trapping frequency thereby increasing 
the single particle excitation energies.  The lowest mode is however qualitatively different: The
 excitation energy first \emph{decreases}  reaching a minimum 
at a "critical" two-body binding energy $\epsilon_b^{c}$ (we will 
justify this name shortly), after which it increases for  stronger attraction. 
This non-monotonic behaviour cannot be understood from a  single-particle picture. Instead, it is due to pair correlations.
The energy cost of exciting a pair of time-reversed states across the energy gap,  as illustrated in Fig.\ \ref{TimereverseFig}(a)-(b), 
 initially decreases with increasing attraction, since the two 
excited particles can use the available  states in the empty shell to increase their overlap. 
In Fig.\ \ref{SpectrumMagicFig}, we normalise   $\epsilon_b$ by 
$\epsilon_b^{c}$, defined as the two-body binding energy which gives the minimum monopole excitation energy, so that  
 we can compare results  for different particle numbers and for the thermodynamic limit. 
Exact values of $\epsilon_b^{c}$ are given in the  Supplementary Material~\cite{Note1}. 


To link the few-body spectrum to the thermodynamic limit, 
we also plot in Fig.\ \ref{SpectrumMagicFig} the lowest monopole mode  obtained 
from a many-body calculation, which includes fluctuations around the Bardeen-Cooper-Schrieffer (BCS) solution~\cite{BruunHiggs2014} 
(see Supplementary Material~\cite{Note1}). Due to the energy 
 gap in the single particle spectrum for a closed-shell configuration, there is a normal to superfluid 
 quantum phase transition at a critical binding energy $\epsilon_b^{c}$. 
The system is in the normal phase for $\epsilon_b<\epsilon_b^c$, and the lowest monopole mode corresponds to 
vibrations in the pairing energy $|\Delta|$ around the  $\Delta=0$ equilibrium value.
The frequency of this mode decreases with increasing attraction and vanishes at $\epsilon_b^{c}$, signalling a quantum phase 
transition to a superfluid phase. In the superfluid phase, the minimum energy is obtained for $|\Delta|>0$, and the Higgs mode 
corresponds to vibrations in $|\Delta|$ around this minimum. Its energy is  
 approximately given by $2|\Delta|$ (The deviation is due to the breaking of particle-hole symmetry), increasing from zero at the critical point.
When $|\Delta|\ll\hbar\omega$, the Cooper pairs are predominantly formed by  time-reversed 
states in the same shell~\cite{BruunHiggs2014}. 
 Importantly, the Higgs mode is \emph{undamped} in this  regime  due to the discrete nature of the trap level spectrum, 
 which is in sharp contrast  to the other table top  systems mentioned above, where the damping is significant.

Comparing the 3+3 and the many-body spectrum  in Fig.\ \ref{SpectrumMagicFig} clearly shows that 
 the lowest monopole mode for the 3+3 system becomes the few-body precursor 
of the Higgs mode with increasing attraction. The non-monotonic behaviour of its energy  is the smooth few-body analogue of the sharp thermodynamic normal 
to superfluid quantum phase transition with a vanishing Higgs mode frequency at the critical point. 
We also show in 
Fig.\ \ref{SpectrumMagicFig} the lowest monopole mode for the   6+6 system corresponding to a closed-shell configuration with the three lowest  shells  filled.  
The non-monotonic behaviour of the lowest excitation energy is now even more pronounced with a deeper minimum, 
reflecting the gradual few- to many-body transition with increasing particle number. 
   
In the Supplementary Material~\cite{Note1}, we illustrate further the few- to many-body transition  by calculating the spectrum for the  closed shell configurations
 up to 15+15 particles. Since it is numerically intractable to perform exact diagonalisations of Eq.\ (\ref{Hamiltonian})
  beyond 6+6 particles, we use a simplified model, which includes only the highest filled and the lowest two empty shells. 
  This calculation clearly shows a pronounced deepening of the minimum of the excitation energy with increasing particle number.

\begin{figure}
\includegraphics[width=0.8\columnwidth]{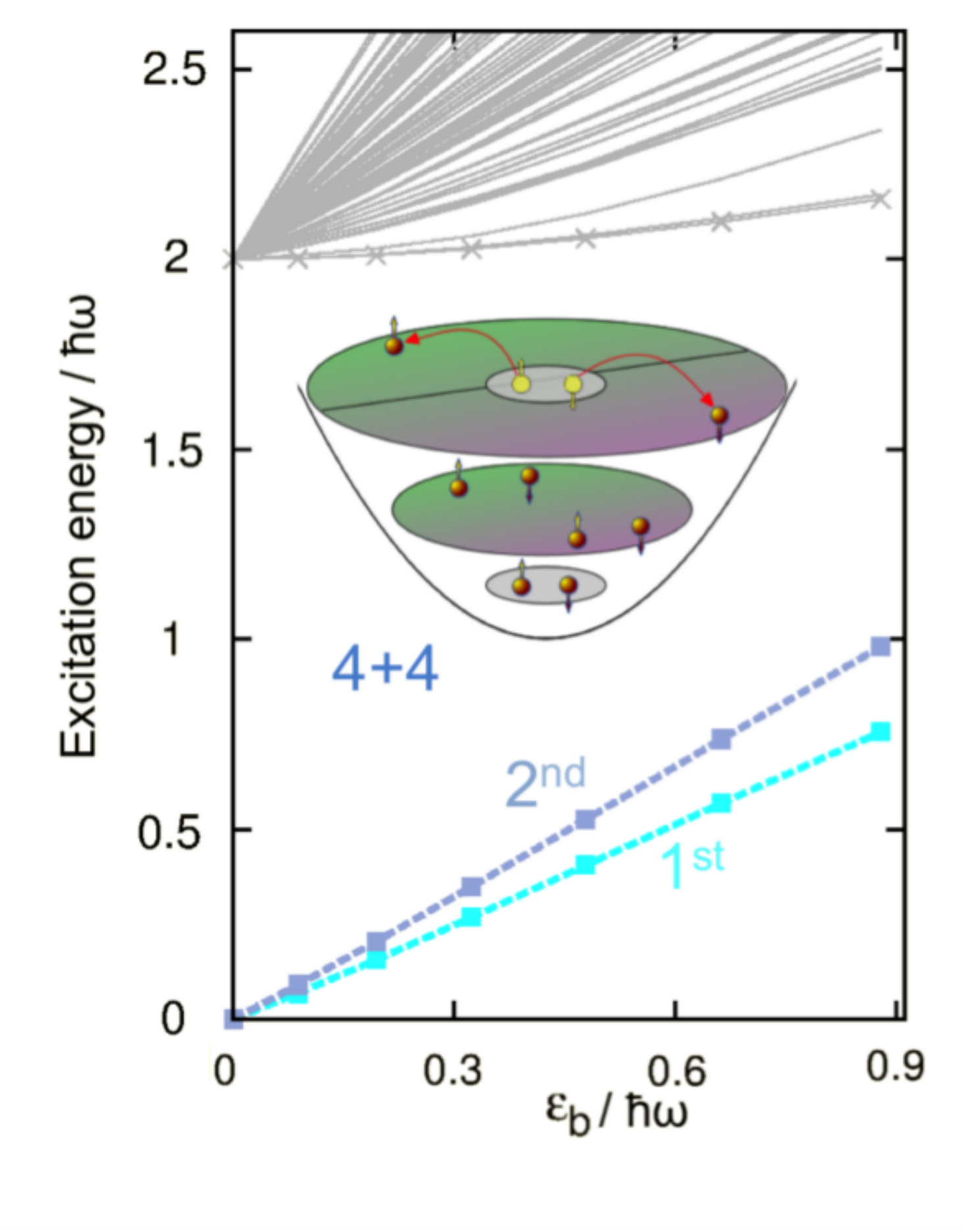}
\caption{\label{SpectrumOpenFig}Monopole excitations of an open-shell system.
 The lowest excitations  are  intra-shell excitations, which do not exhibit a minimum. The
grey lines show higher excitations (which  correspond to inter-shell transitions).  \emph{Inset:} Sketch of time-reversed intra-shell pair-excitations.} 
\end{figure}

In Fig.\ \ref{SpectrumOpenFig}, we plot the lowest monopole excitations for a $4+4$ system, which corresponds to an open-shell configuration 
where there is a pair of $\uparrow\downarrow$ particles in the third shell. Contrary to the closed-shell configuration, 
all excitation energies now increase monotonically with the 
attraction. This is because there is pairing for \emph{any} attractive interaction  so that the lowest excitations involve pair breaking, 
and it demonstrates that the non-monotonic behaviour of the lowest mode energy is characteristic  of a closed-shell configuration, where there is 
a quantum phase transition in the thermodynamic limit.

\begin{figure}
\includegraphics[width=0.5\columnwidth]{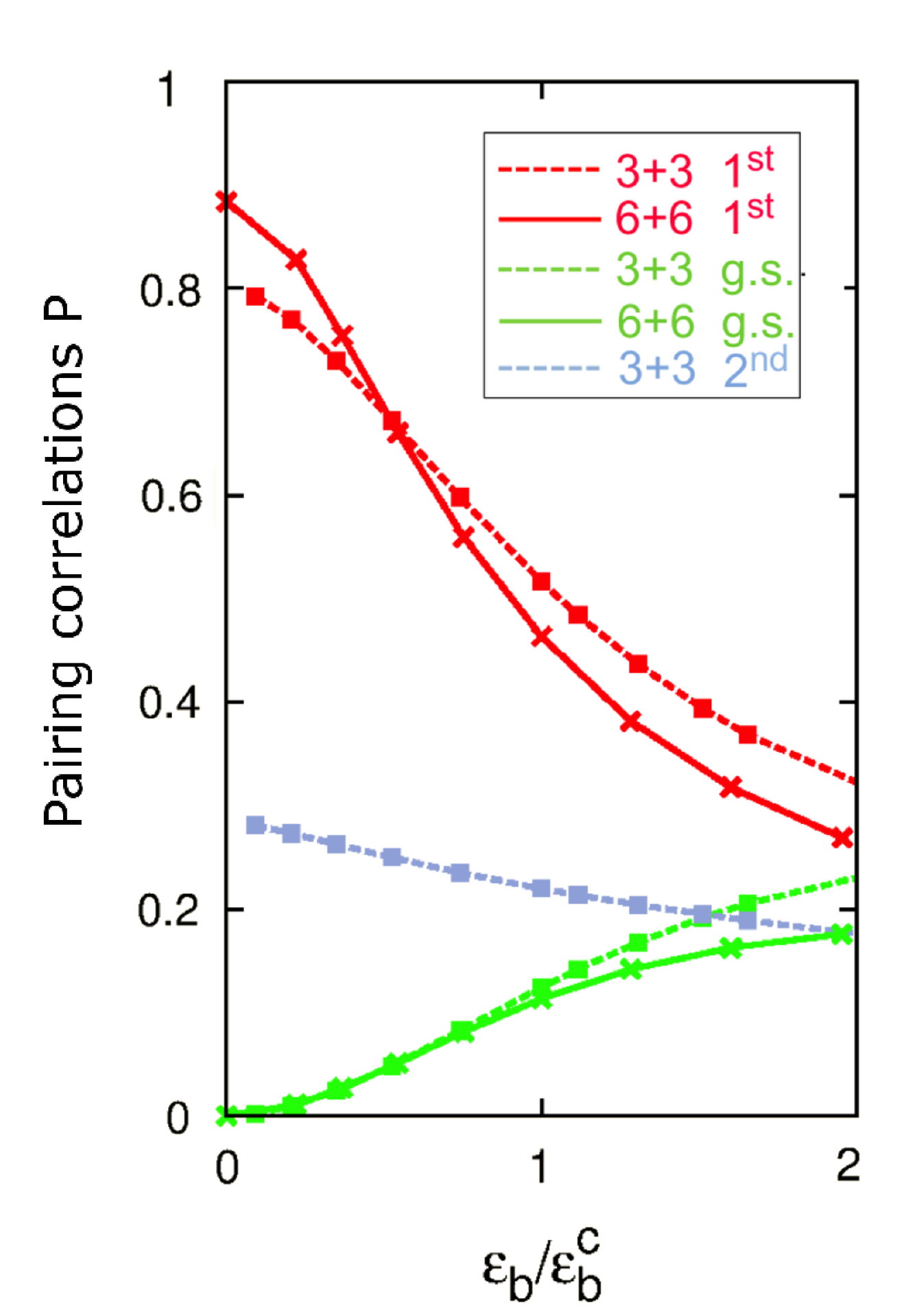}
\caption{\label{CoeffsFig} Pairing  correlations of the few-body states as defined by Eq.\ (\ref{PairCorr}) for  $N=3+3$ fermions (dashed lines) and $N=6+6$ fermions (solid lines).  
The green lines show the ground state and the red lines show the first excited state. The blue line shows the second excited state (only for $3+3$). 
}
\end{figure}

In order  to  investigate further the connection between the few- and many-body physics, we   
quantify the amount of time-reversed pairing correlations in a given state by  
\begin{equation}
P= \sum \limits_{i } \left| C^{tr}_i \right|^2.
\label{PairCorr}
\end{equation}
Here,  $C_i$ are the expansion coefficients in the many-body basis for a given eigenstate. The sum runs over all basis states formed from the 
non-interacting ground state by excitations of time-reversed ("{\it tr}") pairs.   
 In  Fig.\ \ref{CoeffsFig}, we plot $P$ for the ground state and the two lowest  excited states.
Comparing  the first excited state with the ground state and with the second excited states clearly shows that  below the critical binding energy
$\epsilon_b^c$, the wavefunction of the lowest mode is mainly formed by coherent excitations of time-reversed  pairs. 
It is consistent with  the canonical many-body picture of vibrations in $|\Delta|$,
 since such excitations  give rise to fluctuations in the pairing field.
The higher mode has  a significantly smaller proportion of pair correlations, and it  mainly 
consists of  single-particle excitations two shells up. The pairing correlations in the ground state increase with increasing attraction, as it becomes 
more favorable to excite time reversed pairs across the energy gap.
This smooth increase of ground state pair correlations is  the few-body analogue of the normal to superfluid quantum phase transition, 
where excitations of time-reversed pairs cost zero energy  at the critical coupling strength making the system spontaneously form Cooper pairs.  
The pair correlated part of the few-body Higgs mode decreases for $\epsilon_b>\epsilon_b^c$, since 
 it  is orthogonal to the ground state.

\begin{figure}
\includegraphics[height=0.8\columnwidth,angle=90]{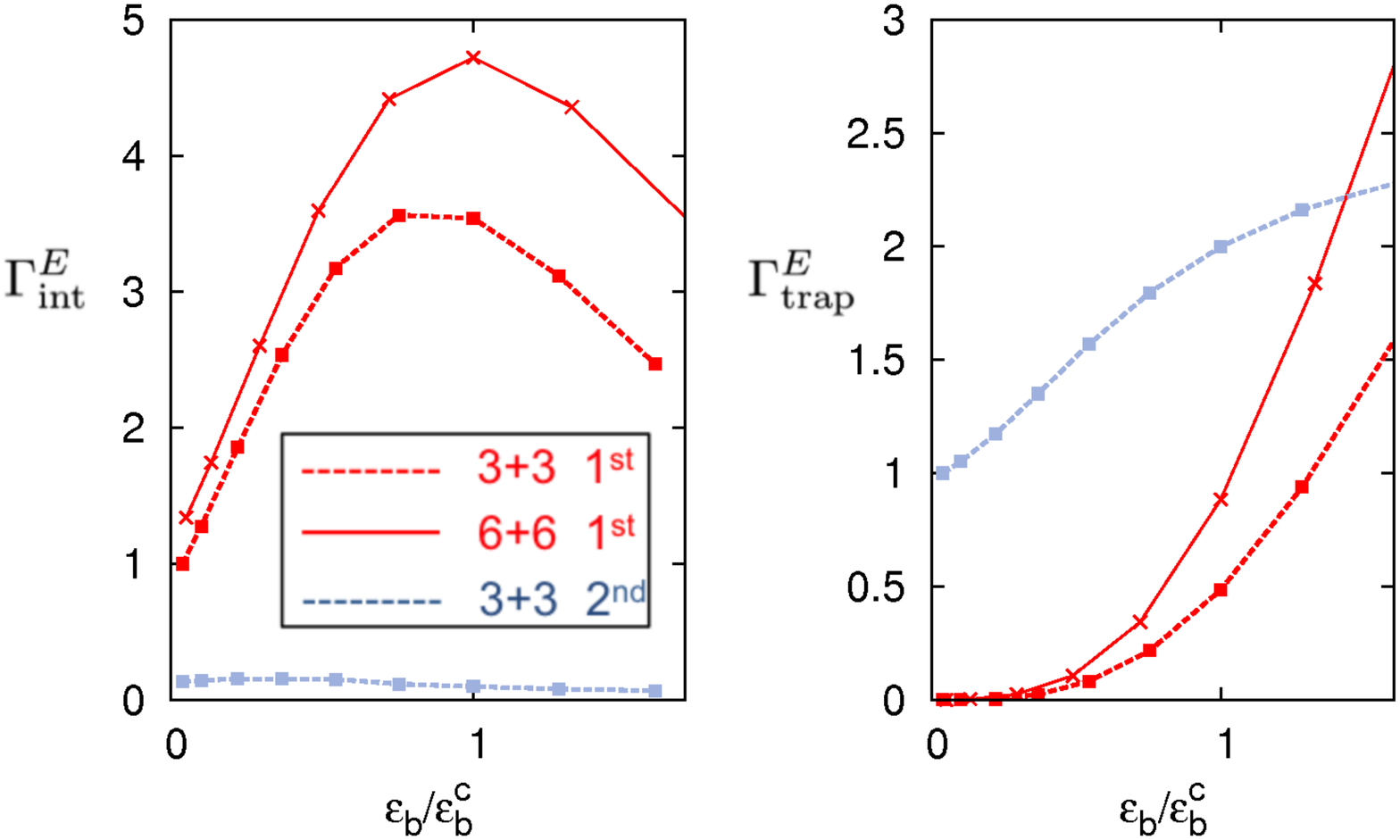}
\caption{\label{xpprobe} Left: Transition matrix elements $\Gamma^E_\text{int}$ corresponding to modulating the interaction strength for a $3+3$ and a $6+6$ system. 
The matrix elements are normalised by  $\Gamma^E_\text{int}$  calculated to the Higgs mode at a very low coupling strength for the $3+3$ system. 
Right: Transition matrix elements $\Gamma^{E}_\text{trap}$ corresponding to modulating the trapping frequency. 
The matrix elements are normalised by  $\Gamma^E_\text{trap}$
  to the second excited state  calculated at 
a very low coupling strength. }
\label{MatrixElementFig}
\end{figure}

We now address how one can detect the few-body Higgs mode in atomic  gas experiments using microtraps.
Two experimental probes are widely 
used: Periodic modulations of the trapping frequency and of the interaction strength. From Fermi's golden rule, the transition rate from 
the ground state $|G\rangle$ to an excited state  $|E\rangle$ is  proportional to the transition 
matrix elements 
\begin{align}
\Gamma^{E}_\text{trap}&=|\langle G|\sum_{i}{\mathbf r}_i^2|E\rangle|^2 \nonumber\\
\Gamma^E_\text{int}&=|\langle G|\sum_{k,l}\delta({\mathbf r}_k-{\mathbf r}_l)|E\rangle|^2
\label{Probes}
\end{align}
for the two probes. 
In Fig.\ \ref{MatrixElementFig}, we plot  $\Gamma^{E}_\text{trap}$ and $\Gamma^{E}_\text{int}$
  to the excited states of the $3+3$ and the $6+6$ systems. Figure \ref{MatrixElementFig} (left) shows that  
the transition rate  into the lowest mode is much larger than the rate 
into the second excited state when the coupling strength is modulated. 
 This is because the interaction operator $\Gamma_\text{int}$ can excite time-reversed pairs, 
(see Supplementary Material~\cite{Note1}), which are precisely the excitations that give rise to  pair vibrations. 
Thus, the Higgs mode can be selectively excited by modulating the interaction strength, using for instance a Feshbach resonance. 
 This fact, together with the non-monotonic frequency behaviour,  can   be used to experimentally identify the Higgs mode. 
On the other hand, 
Fig.\ \ref{MatrixElementFig} (right) shows that when the trapping potential is modulated, the transition rate into the
 second excited state is much larger than into the lowest mode for small attraction. 
The reason is that $\sum_{i}{\mathbf r}_i^2$ is a single particle operator, whereas the lowest mode mostly consists of   time-reversed pair excitations. 
With increasing attraction, the transition rate into the lowest mode increases, consistent with 
the fact that the pair correlation $P$  in the Higgs mode decreases with increasing coupling.

In conclusion, we demonstrated using exact diagonalisation  that the lowest monopole excitation energy of a two-component Fermi gas  
exhibits a non-monotonic behaviour with increasing attractive interaction for closed shell configurations. The mode frequency 
 has a minimum in a cross-over region, which deepens 
  as the many-body limit is approached  with increasing particle number. Comparing with a many-body calculation, 
  we identified the few-body precursor of the Higgs mode,  which has a vanishing frequency at the quantum phase transition
   point between a normal and a superfluid phase. We showed that the mode is mainly formed by coherent excitations of time-reversed 
   pairs, and that it can be selectively excited by modulating the interaction strength. These results demonstrate 
   how a new generation of cold atom experiments using microtraps can be used to explore  
two fundamental questions in physics: The nature of the Higgs mode and the cross-over from few- to many-body physics. 
Our results are also relevant to the nuclear structure community, since we show how cold atoms can be used to  probe 
  pair correlations  in a finite systems much more systematically compared to what is possible in  nuclei~\cite{Frauendorf:2014kq,Potel2013}.

We end by noting that similar results hold for atoms in a 3D trap~\cite{Bruun2001,Bruun2002}. Focus was here on
  the 2D case, as it is closer to being experimentally realised. Indeed, 
  the first experiment observing pairing correlations in 2D has already been reported~\cite{Ries2015}.

\section{Acknowledgements}

We thank Ben Mottelson and Sven {\AA}berg for many useful discussions, as well as Jeremy Armstrong  for a comparison to a more phenomenological pairing model. 
We thank Massimo Rontani for discussions regarding the regularization scheme, and also acknowledge discussions with Selim
 Jochim and Frank Deuretzbacher.  This research was financially supported by the Swedish Research Council and  NanoLund at Lund University.
GMB would like to acknowledge the support of the Villum Foundation via grant VKR023163 and  ESF POLATOM network.

%


\newpage
\hspace{1cm}
\newpage

\section{Supplemental Material for "Few-body precursor of the Higgs mode in a  Fermi gas "}

\subsection{Many-body theory}
To calculate the collective mode spectrum in the many-body limit, we use a BCS mean-field approach combined with Gaussian fluctuation theory. 
When the trap level spacing is larger than the pairing energy, the Cooper pairs are predominantly formed by \emph{intrashell} correlations 
 between time-reversed states in the same shell, i.e.\ $(n,m,\uparrow)$ and $(n,-m,\downarrow)$~\cite{Bruun2002b,Heiselberg2002}. Using this and neglecting the weak angular momentum 
 dependence of the pairing, the mean-field Bogoliubov-de Gennes 
 equations can be reduced to the gap equation~\cite{BruunHiggs2014}
 \begin{equation}
\sum_{n}\frac{1}{2\sqrt{\xi_n^2+\Delta_n^2}}=\sum_{n}\frac{1}{2\epsilon_n-E_2}.
\label{GapEqn2}
\end{equation}
Here, $\Delta_n=\Delta/\sqrt{n+1}$ is the gap for shell $n$, $\epsilon_n=(n+1)\hbar\omega$, and 
$\xi_n=\epsilon_n-\epsilon_F$. The Fermi energy $\epsilon_F=(n_F+3/2)\hbar\omega$ is  between the highest occupied $n_F$
and the lowest unoccupied  shell $n_F+1$. 
Since there is a gap in the single particle spectrum for a closed shell configuration, there is superfluid pairing only above 
a  critical binding energy $\epsilon_b^{c}$. 

For small pairing energy, we can expand (\ref{GapEqn2})  in $\Delta_n/\hbar\omega$ and 
$\epsilon_b/\hbar\omega$. This yields 
 \begin{equation}
 \frac{\epsilon^c_b}{\omega}=\frac{B(n_F)}{2\xi(2)}[\sqrt{1+4\xi(2)/B( n_F)^2}-1]
 \label{ClosedshellCritical}
 \end{equation}
for the critical attraction strength for pairing  with $B(n_F)=\gamma+4\ln 2+\ln n_F$. Here $\xi(z)$ is Riemann's zeta function
and 	$\gamma=0.577$ is the Euler-Mascheroni constant. 
For $\epsilon_b>\epsilon_b^c$, this expansion yields the approximate solution to the gap equation  
\begin{equation}
 \Delta_{n_F}=\frac{\hbar\omega}{\sqrt{7\xi(3)}}\sqrt{\frac{\hbar\omega}{\epsilon^c_b}-\frac{\hbar\omega}{\epsilon_b}+
 \xi(2)\left(\frac{\epsilon_b}{\hbar\omega}-\frac{\epsilon^c_b}{\hbar\omega}\right)}.
 \label{ClosedshellGap}
 \end{equation}

To describe the collective modes, we include Gaussian fluctuations of the pairing field around the mean-field BCS solution.  These  can for low energy 
be split in to phase fluctuations corresponding to Goldstone modes, and amplitude fluctuations corresponding to the Higgs mode. The equation 
determining the Higgs mode energy $\hbar\omega_H$ reads

\begin{equation}
 \sum_n\frac{2\xi_n^2}{E_n(4\xi_n^2+4\Delta_n^2-\hbar^2\omega_H^2)}=\sum_{n}\frac{1}{2\epsilon_n-E_2}.
 \label{HiggsMode}
 \end{equation} 
Assuming perfect particle-hole symmetry around the Fermi level,  we see from the gap equation (\ref{GapEqn2})
 that  $\omega=2\Delta_{n_F}$ is a solution. 
 In the normal phase for $\epsilon_b<\epsilon_b^c$, (\ref{HiggsMode})  has to be solved with 
$E_n=|\xi_n|$ and the   amplitude modes correspond to coherently either 
 adding or  removing a pair of particles.
 The particle-conserving collective modes correspond to subsequently adding and removing a pair of particles, and their 
 frequencies are therefore twice the frequency  obtained by solving (\ref{HiggsMode}).
   Close to the critical coupling strength   $\epsilon_b\lesssim\epsilon_b^c$,
we expand (\ref{HiggsMode}) in $\epsilon_b/\hbar\omega$ and $\epsilon/\hbar\omega$ arriving at 
\begin{equation}
 \frac{\omega_H}{\omega}=\frac 2{\sqrt{7\xi(3)}}\sqrt{\frac{\hbar\omega}{\epsilon_b}-\frac{\hbar\omega}{\epsilon^c_b}+
 \xi(2)\left(\frac{\epsilon^c_b}{\hbar\omega}-\frac{\epsilon_b}{\hbar\omega}\right)}.
 \label{HiggsModeNormalClose}
 \end{equation}

  The solid lines in Fig.\ 1 in the main text are obtained from a numerical solution of (\ref{HiggsMode}) for $n_F=20$, and the dashed lines are
  obtained from $2\Delta_{n_F}$ using (\ref{ClosedshellGap}) for $\epsilon_b>\epsilon_b^c$, and from (\ref{HiggsModeNormalClose}) for $\epsilon_b<\epsilon_b^c$.

 \subsection{Interaction and excitation of time-reversed pairs} 
The interaction term in the Hamiltonian reads in second quantisation 
\begin{gather}
H_{\text{int}}                    \simeq
    g\sum_{\substack{nm\\n'm'}}     \langle nm,n-m|\delta(\mathbf{r}_1-\mathbf{r}_2)|n'm',n'-m'\rangle \nonumber\\               
                    \times a_{nm\uparrow}^\dagger a_{n-m\downarrow}^\dagger a_{n'-m'\downarrow} a_{n'm'\uparrow},           
\end{gather}
where $a_{n_1m_1\sigma}$ removes a particle with harmonic quantum numbers $(n_1,m_1)$ and spin $\sigma$. We  
 have assumed that there are only pair correlations between time reversed states. 

Neglecting the weak angular momentum dependence of the matrix element, we can write the interaction in the form 
\begin{equation}
 H_{\rm int}=G\sum_{nn'}\Gamma_n^\dagger \Gamma_{n'}
\end{equation}
where  $\Gamma_n^\dagger=\sum_{m}a_{nm\uparrow}^\dagger a_{n-m\downarrow}^\dagger/\sqrt{n+1}$. 
The effective coupling strength is 
\begin{equation}
G =\frac{2\pi\int_0^\infty rdr\rho_{n_F}^2(r)}{n_F+1}
\end{equation}
where $\rho_n(r)$ is the radial density from a full $n$-shell. We see that this interaction precisely excites time-reversed pairs across the energy gap. 
Modulating the coupling strength $G$ will therefore strongly couple to the Higgs mode.

\subsection{Few- to many-body transition in a coreless 2D Harmonic Oscillator}
We shall in this section explore  the few- to many-body transition further, by calculating the collective mode 
spectrum for larger particle numbers. Unfortunately,  
the complexity of the many-body problem makes an exact solution via the configuration interaction (CI) diagonalization method described in the main article numerically intractable
 for more than 6+6 particles. We therefore turn to an approximate model that nevertheless contains the relevant physics.   
In the main article, we establish that the formation of the Higgs mode is associated with time-reversed pair excitations from the uppermost filled shell into 
higher empty shells. The energy of the lowest monopole mode initially decreases with increasing attraction, since the pairs can use the degeneracy of the empty 
shells to increase their spatial overlap. This effect becomes more pronounced for larger systems with a larger degeneracy of the empty shells. 
One example of this is the lower minimum in the Higgs excitation energy for the 6+6 system relative to the 3+3 system.

\begin{center}

\begin{figure}[h]
 \centering
 \includegraphics[scale=1.65]{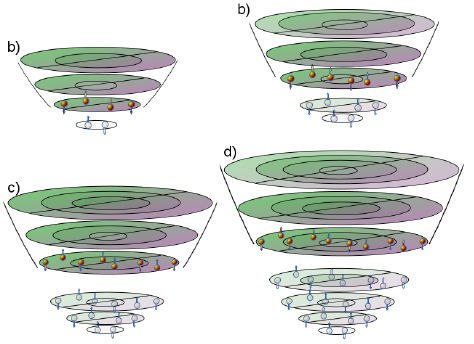}
 \caption{Sketch of the three-level model, where the transparent shells and particles represent the core part of the system, which plays no role 
 in the dynamics. The system with 3+3 particles is approximated by a 2+2 three-shell system (a),  6+6 particles by a three-shell 3+3 system (b),  10+10 particles
 by a three-shell 4+4 system (c), and  15+15 particles by a three-shell 5+5 system (d).}
 \label{fig:1}
\end{figure}

\end{center}

To describe this effect in a simplified numerically tractable system, 
we  consider here a three-level model, where the dynamics of the filled low-lying core of closed shells is ignored assuming that it remains completely filled, see the sketch in Fig.\ \ref{fig:1}. 
The model includes the three most important single-particle harmonic oscillator shells for the low lying  collective modes:  the highest filled and 
the  two lowest empty shells for a given closed shell configuration.  For example, 
 we   approximate the 6+6 system by a 
  three-shell system loaded with 3+3 particles.   This simplification allows us to go to much larger 
particle numbers  than what is possible within the full CI scheme. 

\begin{figure}[h!]
 \centering
 \begin{center}
  
 \includegraphics[scale=0.44,keepaspectratio=true]{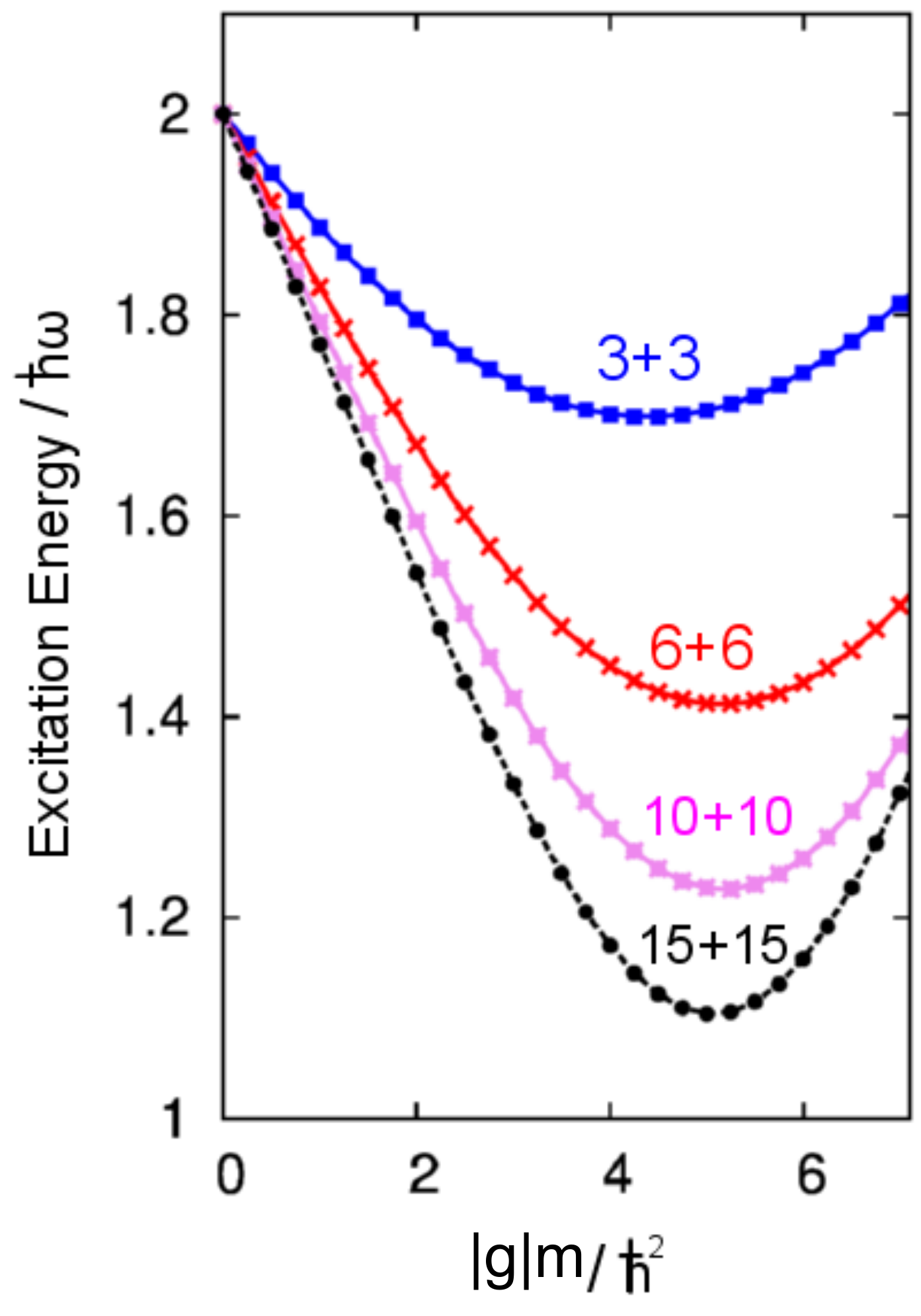}
 \caption{The lowest monopole excitations of the three-shell model for  3+3 (blue line), 6+6 (red line), 10+10 (magenta line) and 15+15 (black line) particles.
  The results are obtained by exact diagonalization of the full interacting many-body Hamiltonian of the three-shell systems.}
 
 \label{fig:2}

 \end{center}

 \end{figure}

In Fig.\ \ref{fig:2}, we plot the lowest monopole excitation energy as a function of the coupling strength $|g|m/\hbar^2$ 
of the three-level model for the closed shell configurations up to 15+15 particles.
We see that the minimum of the mode energy deepens with increasing particle number, reflecting that 
 the number of degenerate pair excitations increases, which allows for a larger spatial overlap between particles in the upper shells. 
 This clearly demonstrates how  with increasing particle number, the finite size system gradually approaches the many-body limit, where 
 the Higgs mode energy vanishes at the critical coupling strength and it costs zero energy to coherently excite pairs across the shells. 
Note that the excitation energy minimum for the 3+3 system is deeper for the three-level model as compared to that of the full model shown in Fig.\ 1 in the main article. 
This is because coherent excitations of time-reversed pairs play a larger role in the three-level model, which has a much smaller phase space as 
compared to the full model. 

%
 
%

\section{Numerical Calculations}
We use two parameters which together define the many-body basis used in our calculations:  the 
single particle energy cut-off $E^\text{shell}_\text{max}$, and the many-body cut-off $E_\text{cut}$.
These two cutoff parameters are optimized  for each calculation individually in order  to reach the best possible convergence.
For the transition matrix elements, the comparison between the two different system sizes is more delicate, and we therefore 
follow a slightly different strategy: The same $E^\text{shell}_\text{max}$ is used for the  $3+3$ and $6+6$ systems,
and  $E_\text{cut}$ is defined  so that the maximum many-body excitation energy
relative the non-interacting ground state is the same for both systems.
In this way, we have a systematic way of comparing matrix elements for the two different system sizes.

\subsection{Critical binding energy}
In the main manuscript, we divide  the two-body binding energy $\epsilon_b$ by the critical binding energy $\epsilon_b^c$ when  the  excitation energies
are plotted, so that 
 we can compare results obtained for different system sizes. The lowest monopole mode will then by definition have the minimum energy at $\epsilon_b/\epsilon_b^c=1$ for 
all system sizes. Of course, the actual value of $\epsilon_b^c$ depends on the system size. The exact calculations give
  $\epsilon_b^c=0.86\hbar\omega$ 
for the 3+3 system and   $\epsilon_b^c=0.78\hbar\omega$ for the 6+6 system. This is significantly larger than what is predicted by the many-body expression 
Eq.\ (\ref{ClosedshellCritical}) for $n_F=2$ and $n_F=3$.
This is not surprising as the systems are  small and far from the thermodynamic limit, and since the microscopic model includes all excitations, 
whereas the many-body theory focuses on the pair excitations.

\end{document}